\documentclass[aps,preprint,draft,showpacs,floatfix]{revtex4}

\usepackage{longtable}

\usepackage[final]{graphicx}
\usepackage{dcolumn}
\usepackage{bm}
\newcommand{\mm}[1]     {\ifmmode {#1} \else{}${#1}$\fi}
\newcommand{\mmm}[1]    {\ifmmode{}#1 \else{}${#1}$\fi}
\newcommand{\beq}[1]{\begin{equation}\label{#1}}
\newcommand{\eeq}{\end{equation}}

\def \cacuni{\mm{\rm Ca_{3}Cu_{3-x}Ni_x(PO_4)_4}}
\def \acu{\mm{\rm A_{3}Cu_{3}(PO_4)_4}}

\def \l{$\lambda$}

\def\vec#1{\mm{{\rm\bm{{\mathrm#1}}}}}

\begin{document}
			

\title{\large Crystal and magnetic structures of the spin-trimer
  compounds $\bm\cacuni$ (x=0,1,2) }

\author{V.~Yu.~Pomjakushin}
\affiliation{Laboratory for Neutron Scattering, ETH Zurich and Paul Scherrer
Institut, CH-5232
Villigen PSI, Switzerland}
\author{A.~Furrer}
\affiliation{Laboratory for Neutron Scattering, ETH Zurich and Paul Scherrer
Institut, CH-5232
Villigen PSI, Switzerland}
\author{D.~V.~Sheptyakov}
\affiliation{Laboratory for Neutron Scattering, ETH Zurich and Paul Scherrer
Institut, CH-5232
Villigen PSI, Switzerland}
\author{E.~V.~Pomjakushina}
\affiliation{Laboratory for Developments and Methods, PSI and
  Laboratory for Neutron Scattering, ETH Zurich and Paul Scherrer
  Institut, CH-5232 Villigen PSI, Switzerland}
\author{K.~Conder}
\affiliation{Laboratory for Developments and Methods, PSI, CH-5232
Villigen PSI, Switzerland}

\date{\today}

\begin{abstract}

Crystal and magnetic structures of a series of novel quantum spin
trimer system \cacuni\ (x=0,1,2) were studied by neutron powder
diffraction at the temperatures 1.5-290~K. The composition with one
Ni per trimer (x=1) has a monoclinic structure (space
group $P2_1/a$, no.~14) with the unit cell parameters $a=17.71$~\AA,
$b=4.89$~\AA, $c=8.85$~\AA\ and $\beta=123.84^\circ$ at T=290~K. The
(x=2) composition crystallizes in the $C2/c$ space group (no.~15)
with the doubled unit cell along $c$-axis. Each trimer is formed by
two crystallographic positions: one in the middle and the second one
at the ends of the trimer.  We have found that the middle position is
occupied by the $\rm Cu^{2+}$, whereas the end positions are
equally populated with the $\rm Cu^{2+}$ and $\rm Ni^{2+}$ for (x=1)
while in the (x=2) the trimers were found to be of only one type
Ni-Cu-Ni. Below $T_N=20$~K the (x=2) compound shows an
antiferromagnetic ordering with propagation vector star
$\{[{1\over2},{1\over2},0],[-{1\over2},{1\over2},0]\}$. The magnetic
structure is very well described with the irreducible representation
$\tau_2$ using both arms of the star $\{\vec{k}\}$ with the magnetic
moments $1.89(1)\,\mu_B$ and $0.62(2)\,\mu_B$ for $\rm Ni^{2+}$ and
$\rm Cu^{2+}$ ions, respectively. We note that our powder
diffraction data cannot be fitted by a model containing only one arm
of the propagation vector star. The Cu/Ni-spins form both parallel
and antiparallel configurations in the different trimers, implying
substantial effect of the inter-trimer interaction on the overall
magnetic structure.

\end{abstract}


\maketitle


\section{Introduction}

The low-dimensional magnets have been attracting attention during
last years since they show new interesting quantum effects and also
because they are considered as a model system to study very complex
phenomena, like high-temperature superconductivity in metal oxides.
\acu\ (A=Ca, Sr, Pb) is a novel quantum spin trimer system
\cite{belik02} in which the three ${\rm Cu^{2+}} (S={1\over2})$
spins are antiferromagnetically coupled giving rise to a doublet
ground state, as determined by neutron spectroscopy
\cite{matsuda05}. The trimer clusters form (1D) chains with weak but
not negligible intertrimer interaction \cite{belik05, drillon93}
leading to a long-range magnetic ordering at $T_C=0.91$~K,
$T_N=0.91$~K and $T_N=1.26$~K for A = Ca, Sr and Pb, respectively.
By substituting a $\rm Cu^{2+}$ spin in the trimer by $\rm Ni^{2+}$
$(S=1)$ a singlet ground state could be in principle realized
offering the observation of the Bose-Einstein condensation (BEC) in
a quantum spin trimer system similar to the field-induced BEC of the
bosonic triplet state in the spin dimer system $\rm TlCuCl_3$
observed by the inelastic neutron scattering \cite{ruegg03}. The
substitution of Cu by Ni was successfully realized in the \cacuni\
(x=1,2) resulting in the new mixed trimer phase with the structure
parameters close to the ones of the pristine material (x=0). The
magnetic excitations in this series were studied by the inelastic
neutron scattering \cite{furrer07} making use of the structure data
reported in the present paper. The observed excitations were
associated with transitions between the low-lying electronic states
of trimers. The nearest-neighbor exchange interactions within the
trimers in the (x=1,2) compounds were determined to be also
antiferromagnetic with $J_{\rm Cu-Cu}=-4.92(6)$~meV and $J_{\rm
Cu-Ni}=-0.85(10)$~meV and an axial single-ion anisotropy parameter
$D_{\rm Ni}=-0.7(1)$~meV. The ground state was found to be doublet,
triplet and quintet in the Cu-Cu-Cu, Cu-Cu-Ni, and Ni-Cu-Ni trimers,
respectively that are the basic constituents of the title compounds.
The hypothesis of realizing the singlet ground state
that motivated the present work was not met, but
without the detailed structural information the analysis of the
magnetic excitations could not be performed. In this paper we
present the results of the neutron and synchrotron x-ray powder
diffraction study of the crystal and magnetic structures of \cacuni\
(x=0,1,2).

\section{Samples. Experimental}
\label{exp}

Polycrystalline samples of \cacuni\ (x=0,1,2) were synthesized by a
solid state reaction using CuO, NiO, $\rm CaCO_3$ and $\rm
NH_4H_2PO_4$ of a minimum purity of 99.99\%. The respective amounts
of the starting reagents were mixed and heated in alumina crucibles
very slowly up to 600$^\circ$C and then annealed at 900$^\circ$C
during at least 100 h, with several intermediate grindings. The $ac$
magnetic susceptibility $\chi(T)=\chi'(T)+i\chi''(T)$ was measured
in zero external field with amplitude of the $ac$ field 10~Oe and
frequency 1~kHz using Quantum Design PPMS station. Neutron powder
diffraction experiments were carried out at the SINQ spallation
source of Paul Scherrer Institute (Switzerland) using the
high-resolution diffractometer for thermal neutrons HRPT \cite{hrpt}
($\lambda=1.866$~\AA, high intensity mode $\Delta
d/d\geq1.8\cdot10^{-3}$), and the DMC diffractometer \cite{dmc}
situated at a supermirror coated guide for cold neutrons at SINQ
($\lambda=4.2$~\AA). All the temperature scans were carried out on
heating. x-ray synchrotron diffraction measurements at room
temperature were done at the Material Sciences beam line (MS,
SLS/PSI). The refinements of the crystal and magnetic structure
parameters were carried out with {\tt FULLPROF}~\cite{Fullprof}
program, with the use of its internal tables for scattering lengths
and magnetic form factors.

\section{Crystal structure}
\label{cryst_sec}

Both (x=0) and (x=1) compounds have the same space group $P2_1/a$
with the structure parameters shown in Table \ref{tab1}. The
diffraction pattern and the refinement plot for the (x=1) sample is
shown in Fig. \ref{dpcu2ni}. There is a small admixture of two
impurity phases: whitlockite $\rm Ca_{19}Cu_2(PO_4)_{14}$ and nickel
oxide $\rm NiO$ that are indicated by the additional rows of tics.
The structure parameters of the whitelockite phase were fixed by the
values reported in Ref.~\cite{lazoryak99}. The NiO phase has two
rows of tics due to additional antiferromagnetic phase. The mass
fractions of the whitlockite and NiO are 2.2(2)\%, 0\%, and
3.3(1)\%, 0.20(2)\% in the (x=0) and (x=1) compositions,
respectively. An important block of structure is the Cu2-Cu1-Cu2
trimer that is shown in Fig.~\ref{trimer} together with the
surrounding $\rm PO_4$ tetrahedra. The crystal structure consists of
$\rm PO_4$ tetrahedra connecting the Cu trimers in chains running
along $a$-axis as shown in Fig.~\ref{cstr}. Figure~\ref{trimer4}
shows the view of the trimer connectivity projected roughly to the
$(ab)$-plane.

One can see that the strongest intertrimer interaction
is expected to be along $b$-axis giving the quasi (1D) trimer chains
(...AB...) and (...CD...), where A, B, C and D denote the trimers
shown in Fig.~\ref{trimer4}. There are two superexchange paths
between the Cu1(A) and Cu2(B) ions that go through two $\rm
PO_4$-tetrahedra Cu1(A)-O-P1-O-Cu2(B), and also two similar paths
between Cu2(A)-Cu1(B) ions. In addition, the distance between the
Cu2(A)-Cu1(B) ions is the shortest one ($d_{\rm
Cu2(A)-Cu1(B)}=3.4$~\AA, whereas $d_{\rm Cu1(A)-Cu1(B)}=4.9$~\AA)
providing the largest dipole interaction. The interaction between
the 1D chains of trimers along $a$-axis is mediated by the two
superexchange paths between Cu2 ions Cu2(A)-O-P2-O-Cu2(C) as shown
in Fig.~\ref{trimer4}. The intertimer interaction along $c$-axis
(Fig.~\ref{cstr}) is the weakest one since the path contains an
additional Ca-O link and the ions are separated by large distance
$d_{\rm Cu2-Cu2}=6.5$~\AA.

The Ni-atom in the (x=1) sample occupies the Cu2 positions at the
ends of the trimers, whereas the middle position Cu1 is occupied by
Cu. The end positions are equally populated by both Cu and Ni ions.
The occupancy factors can be reliably refined due to significantly
different coherent scattering lengths of Ni (10.3~fm) and Cu
(7.7~fm) nuclei. The fit model assumed that we have two Cu atoms and
one Ni atom per formula unit allowing them to occupy both Cu1 and
Cu2 sites. The refined occupancies are listed in Table~\ref{tab1}.

The composition with two Ni atoms (x=2) crystallizes in a
different space group $C2/c$ with a doubled unit cell along
$c$-axis. The transformation from the $P\,1\,2_1/a\,1$ to the
$C\,1\,2/c\,1$ structure is given by the matrix $\vec{A}=\vec{a}$,
$\vec{B}=\vec{b}$, $\vec{C}=2\vec{c}$ and the origin shift
$\vec{p}=\vec{a}/2$. The structure solution was done with the {\tt
FOX} program \cite{fox} using the synchrotron x-ray diffraction
pattern collected at the wavelength $\lambda=0.9185$~\AA. Final
refinement of the neutron diffraction data resulted in the structure
parameters listed in Table~\ref{tab1}. The experimental and the
refined diffraction patterns are shown in Fig.~\ref{dpcuni2}. In
spite of the doubled unit cell the density of the Bragg peaks is the
same as for the (x=0,1) compositions due to C-centered Bravias
lattice. The mass fractions of the whitlockite and NiO impurities
amounted to 6.4(2)\% and 1.56(3)\%, respectively. The crystal
structure motif in the (x=2) is very similar to the one in the
pristine compound and thus all the crystal structure parameters can
be directly compared (see Table~\ref{tab1}). The noticeable change
in the trimer structure is the decrease in Cu1-O4-Cu2 bond angle.
Similar to the (x=1) compound the Ni-atoms predominantly occupy the
end positions, while the middle position is mainly occupied by Cu.
Table~\ref{bvstab} shows the average cation-oxygen bond lengths and
the distortions of all the polyhedra and the bond valence sum BVS
for all cations calculated from the experimental distances using
{\tt FULLPROF} \cite{Fullprof} suite and the BVS parameters from
Ref.~\cite{brown85}. The BVS for the oxygen atoms (not shown in the
table) are very close to 2.

\section{Magnetic structure of ${x=2}$ compound}
\label{mag_str}

The magnetic susceptibility data are presented in Fig. \ref{X(T)}.
The susceptibility of the composition with (x=1) that does not
exhibit magnetic ordering down to 1.5~K is also shown in the plot
for comparison. The high-temperature part of the susceptibility
(T=75-225~K) was fitted to $\chi(T)=C/(T-T_{\rm CW}) +B$, where $C=N_{\rm A}
2S(S+1)\mu_{\rm B}/3k$ is the Curie constant, $B$ is a constant background term due to the
impurity phases. The fit results are shown in the inset of Fig.
\ref{X(T)}. The paramagnetic spin values per magnetic site $\rm Cu^{2+}$/$\rm Ni^{2+}$
calculated from the refined values of the Curie constant C amounted to $S=0.92$(1) and 0.96(2) that are in reasonable agreement with the
expected average spin-values per magnetic site 0.67 and 0.83 for x=1 and x=2 compositions,
respectively.  The broad peak at $T_N=20$~K is associated with a
transition to the magnetically ordered state. The high statistics
neutron diffraction (ND) patterns were collected at the temperatures
below ($T=1.5$~K) and above (25~K) the transition at $T_N$ using
neutron wavelength \l=4.2~\AA. The low temperature pattern possesses
many additional magnetic Bragg peaks that prove the presence of long
range magnetic ordering. The difference pattern (``1.5~K''-``25~K'')
containing purely magnetic contribution is shown in
Fig.~\ref{difpatt}. The temperature scan performed in the
temperature range from 1.5 to 21~K confirms that the magnetic Bragg
peaks disappear above 18~K.  The Bragg peaks of the difference
pattern can be excellently indexed in the chemical cell of the $x=2$
compound with the propagation vector $\vec{k}=[{1\over2} {1\over2}
0]$, thus proving the antiferromagnetic nature of the transition at
$T_N$. The powder profile matching refinement (Fig.~\ref{difpatt})
shows that all the peaks are well described in the above model
($R_{wp}=8.9$, $R_{\rm exp}=6.7$, $\chi^2=1.73$, $\chi_{\rm
Bragg}^2=1.92$). The small peculiarities near $2\theta\simeq33.15^o$, $81.6^o$,
and $69.4^o$ are due to the slight position mismatch of the intense
nuclear Bragg peaks for the two temperatures.

\subsection{Symmetry analysis}

Using the determined propagation vector we performed the symmetry
analysis according to Izyumov and Naish \cite{izyumov91} to derive
possible magnetic configurations for both Ni (8f) and Cu (4b)
magnetic sites of the space group $C\,1\,2/c\,1$ (no. 15). For this
purpose we used program $\tt BASIREP$ \cite{Fullprof} to obtain
corresponding basis functions $\vec{\psi}_{j0}$ ($3m$-dimensional
vectors) in the $0^{\rm th}$ unit cell of all atoms of the site
$(j)$ with multiplicity $m$. The magnetic moments are obtained by
the linear combination of the basis functions:

\beq{Sofbf}
\vec{S}_{j0}=\sum_{\lambda,k_L}C_{\lambda,\vec{k}_L}\vec{\psi}_{j0},
\eeq

where $\vec{S}_{j0}$ is a $m$-dimensional column of spins on the
position $(j)$, $C_{\lambda,\vec{k}_L}$ are arbitrary mixing
coefficients, $\vec{k}_L$ enumerates the arms of the propagation vector
star \{$\vec{k}$\}, $\lambda$ runs over the basis functions that
appear in the decomposition of the magnetic representation. The
magnetic moments of the atoms displaced by the translation $\vec{t}$
are obtained by the relation:

\beq{Soft}
\vec{S}_j(\vec{t})=\vec{S}_{j0}\exp(2\pi i \vec{k}\vec{t}),
\eeq

where $\vec{t}$ is a centering translation $({1\over2} {1\over2} 0)+$
or chemical cell translations.

The little group  $G_k$ of propagation vector contains two elements
\{$1,\bar{1}$\}. The star of propagation vector has two arms
$\vec{k}_1=[{1\over2} {1\over2} 0]$ and $\vec{k}_2=[-{1\over2} {1\over2}
0]$ that are related by the $2_y$ symmetry operator. The reciprocal
$(\vec{a^*b^*})$-plane showing the propagation vector star in both
centered and primitive unit cells is presented in Fig.~\ref{rec}. There
exist two one-dimensional real irreducible representations $\tau_1$ and
$\tau_2$ with the characters (1,1) and (1,-1), respectively. For
completeness we give the k-vector in the settings used in the Kovalev's
book \cite{kovalev}. Kovalev uses $B\,1\,1\,2/b$ settings for the space
group no.~15 with the transformation matrix: $\vec{A}=\vec{a}$,
$\vec{B}=\vec{c}$ and $\vec{C}=-\vec{b}$, where the capital and
lowercase letters are the basis vectors for $B\,1\,1\,2/b$ and standard
$C\,1\,2/c\,1$ settings. The relationship between the centered $B\,2/b$
cell and Kovalev's choice of primitive lattice is given by (page 57 in
\cite{kovalev}): $\vec{a}=\vec{A}$, $\vec{b}=-(\vec{A}+\vec{C})/2$ and
$\vec{c}=(-\vec{A}+\vec{C})/2$, where the lowercase letters stand for
the primitive cell. Using the above matrices we find that the k-vector
star in the primitive Kovalev basis reads: \{$[0,0,{1\over2}]$,
$[0,{1\over2},0]$\}, which corresponds to the star $\vec{k}_4$ in the
Kovalev's notations. The representation of this $\vec{k}_4$-vector group
contains two one-dimensional irreps $\tau_1$ and $\tau_2$ (pages 59 and
229 in \cite{kovalev}) in accordance with the {\tt BASIREP}
calculations.

The decomposition of the axial vector representations
for Cu and Ni sites reads: $3\tau_2$ and $3\tau_1\bigoplus3\tau_2$,
respectively.  Only $\tau_2$ appears in the decomposition for both
magnetic sites and hence we conclude that the magnetic ordering goes
according to the representation $\tau_2$. In the $0^{\rm th}$ cell
there are two Cu-atoms and four Ni-atoms (centering translation
excluded). Both Ni and Cu sites are split into two independent
orbits as shown in Table~\ref{magtab}.  The atoms on the second
orbits are obtained from the first ones by applying a rotation $2_y$
around $(0,y,{1\over4})$.  Table \ref{magtab} shows the basis
functions of $\tau_2$ that will be used below. Actually, in our case
the basis functions for both orbits and both arms of the star
\{$\vec{k}$\} can be chosen to be the same as the functions for the
orbit 1 and the propagation vector $\vec{k}_1$. However we have also
constructed a special case of the basis functions for the orbit 2
shown in Table~\ref{magtab} for the purposes we explain below.

Inside of each Ni-orbit the spins are antiparallel and have the same
magnitude for the $\tau_2$ representation. The trimers are formed by
the two Ni-atoms related by inversion with the Cu in the inversion
center. Since the Cu atoms are in the positions (0 \mm{{1\over2}} 0)
or (0 \mm{{1\over2}} \mm{{1\over2}}) the inversion about Cu moves
the Ni atom out of the $\rm 0^{th}$ cell to the neighboring cell
shifted either along $x$ or $y$-axis (shift along $z$-axis is not
important). This translation reverses the Ni spin according to
(\ref{Soft}) leaving the Ni spins parallel in the trimers.

\subsection{Magnetic structure determination}

The spin components for both Cu and Ni atoms are unrestricted by
symmetry giving in general 12 independent parameters: one Ni spin
and one Cu spin for each orbit. However, we constrain the sizes of
the spins of the atoms to be equal in both orbits, because the
inner-atomic energies generating the atomic spin are much larger
than the inter-atomic exchange interactions. The atom positions were
fixed by the values determined from the HRPT diffraction pattern
measured at 25~K with $\lambda=1.886$~\AA.  We performed a simulated
annealing minimization \cite{kirkpatrick83} of the integrated
intensities of the 46 magnetic Bragg peaks using {\tt FULLPROF}
program for this general model (A) using one arm of \{$\vec{k}$\}.
Finally the result of the simulated annealing search was refined
using usual Rietveld refinement of the powder diffraction pattern.
The best fit (model A) is shown in Fig.~\ref{difpatt}. The fit
quality is not really good if we compare the $\chi^2=4.5$ with the
one obtained in the powder matching fit $\chi^2=1.7$ as shown in
Fig.~\ref{difpatt}. We have to conclude that there is no
satisfactory solution in the model with one arm of the
propagation vector star \{$\vec{k}$\}. However, for completeness we
present the fit results for this model in Table \ref{magmom}.

We have found a real solution considering both arms of the star
\{$\vec{k}$\}, which excellently fits to the experimental data
(model B in Fig.~\ref{difpatt}). At first, we considered only orbit
1 with the propagation vector $\vec{k}_1$, i.e. only half of the
atoms, and obtained an excellent fit. Then we have constricted the
basis functions of the vector $\vec{k}_2$ from the ones for the
vector $\vec{k}_1$ by the using the relation [formula (9.15) of
Ref.~\cite{izyumov91}]:

\beq{karm}
\vec{\psi}(\vec{k}_2,j')= \exp({-2\pi i \vec{k}_2 \vec{a}_p(g_2,j)}) \delta_{g_2} \hat{R}(h_{g_2}) \vec{\psi}(\vec{k}_1,j),
\eeq

where $g_2$ is the symmetry element generating the arm $\vec{k}_2$,
$j$ and $j'$ are the initial atom number and the atom number after
applying $g_2$, $\vec{a}_p(g_2,j)$ is a translation returning the
transformed by $g_2$ atom $j'$ to the $0^{th}$-cell,
$\hat{R}(h_{g_2})$ is the rotation matrix of the operator $g_2$,
$\delta_{g_2}$ is 1/-1 for the proper/improper rotation.  The basis
functions obtained this way are listed in Table~\ref{magtab}.  Using
these basis functions and considering only orbit 2 with the same
mixing coefficients $C_{\lambda,\vec{k}_2}$ of formula (\ref{Sofbf})
as for the orbit 1 we get identical Bragg peak intensities. Thus,
the set of the structure factors is essentially the same for both
orbits provided that the basis functions are related by the
transformation (\ref{karm}) and the propagation vectors by the
matrix $\hat{R}(h_{g_2})$. We note, that it cannot be a general
assertion valid for arbitrary space group since a crystallographic
site can split into the k-star orbits containing different number of
atoms. The transformation (\ref{karm}) moves the atoms from orbit 1
to orbit 2 and rotates the Cu spin by $\pi$ around $y$ axis, while
for the Ni atoms it reflects the spin about $(ac)$ plane. 

\begin{figure}
  \begin{center}
    \includegraphics[width=2cm]{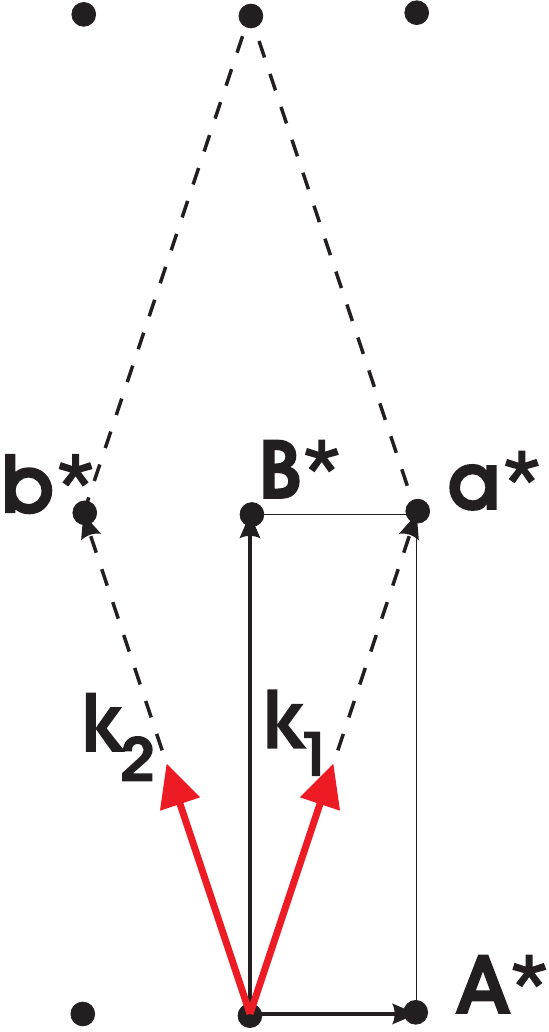} 
  \end{center}

  \caption{The reciprocal $(\vec{a}^*\vec{b}^*)$-plane showing both conventional (direct space centered {$ C\,1\,2/c\,1$} setting) and primitive reciprocal cells by the dotted and solid lines, respectively. The primitive basis vectors are related to the conventional ones as $\vec{a^*}=\vec{A^*}+\vec{B^*}$, $\vec{b^*}=-\vec{A^*}+\vec{B^*}$. The propagation vector star is shown by \{$\vec{k}_1, \vec{k}_2$\}.}
  \label{rec}
\end{figure}

In our case the equivalence of the structure factors for the orbit 1,
$\vec{k}_1$ and the orbit 2, $\vec{k}_2$ with the basis functions listed
in Table~\ref{magtab} can be easily seen if we consider the primitive
unit cell. The primitive basis vectors are related to the ones of the
C-centered lattice as: $\vec{a}=(\vec{A}+\vec{B})/2$,
$\vec{b}=(-\vec{A}+\vec{B})/2$ and $\vec{c}=\vec{C}$, where the
lowercase letters stand for the primitive basis. In the 0th primitive
unit cell, the spins of Cu and Ni on each orbit are parallel and the
spins on different orbits are related by a $\pi$-rotation about \vec{B}.
The propagation vectors are $\vec{k}_1=[{1\over2},0,0]$ and
$\vec{k}_2=[0,{1\over2},0]$ as shown in Fig. \ref{rec}. The origin of
the space group can be shifted by $1\over2$ along \vec{c}-axis. Since
the structure factor is calculated only for the atoms on one orbit we
shift orbit 2 by $(0,0,-{1\over2})$, so that the Cu-atom stays in the same position
$({1\over2},{1\over2},0)$ for both orbits. The Ni11 and Ni12 are in
$(x,y,z)$ and $(1-x,1-y,1-z)$;  the Ni21 and Ni22 are in $(1-y,1-x,z)$
and $(y,x,1-z)$. The magnetic structure factor is
$\vec{F}(\vec{H})\propto\sum_{j}\vec{M}_{\perp{j}}\exp{(\vec{H}\vec{r}_j
)}$, for the scattering vector $\vec{H}=\vec{h}+\vec{k}$, where \vec{h}
is a reciprocal lattice vector of the crystal structure,
$\vec{M}_{\perp{j}}=\vec{H}\times[\vec{M}_j\times\vec{H}]/H^2$ and the sum
runs over Cu and two Ni-atoms at the positions $r_j$ with the magnetic
moments $\vec{M}_j$. The phase factor for Cu is the same for both
orbits, the phase factors for Ni11 and Ni22, and Ni12 and Ni21 will be
the same if we choose the reflections $\vec{H}_1=(h,k,l)+\vec{k}_1$ and
$\vec{H}_2=(k,h,-l)+\vec{k}_2$ for the orbit 1 and 2, respectively. The
reciprocal vectors $\vec{H}_1$ and $\vec{H}_2$ are related by
$\pi$-rotation about \vec{B^*}-axis (Fig.~\ref{rec}) and hence have the
same length and give the Bragg peaks at the same $2\theta$-position in
the powder diffraction pattern. Since the spins on two orbits are also
related by a $\pi$-rotation about \vec{B}-axis, which is collinear to
\vec{B^*}, the vectors  $\vec{M}_{\perp{j}}$ are also related by
a $\pi$-rotation about \vec{B}-axis. Hence, the intensity
$I\propto\mid\vec{F}\mid^2$ will be the same for the Bragg peaks located
at $H_1$ and $H_2$ for orbits 1 and 2, respectively, and the powder
diffraction patterns generated by orbits 1 and 2 will be identical.

The two
orbits do not interfere with each other because of different
propagation vectors. Hence using all the atoms we naturally get the
same fit quality with $\sqrt{2}$ smaller mixing coefficients. The
results of the fit are shown in Fig.~\ref{difpatt} (marked as model
B). This model contains only 6 refinable parameters and gives the
same fit quality as the powder matching refinement, implying that
the fit cannot be better for the given propagation vector $\vec{k}$.
The imperfection of the fits near $2\theta=33.2^\circ$ and
$54.7^\circ$ seen in both the powder match and the model B
difference curves is apparently due to a ``non-ideal'' subtraction
(``1.5~K-25~K'') of the large nuclear peaks (002) and (110) at these
angular positions. Another explanation could be the presence of weak
ferromagnetism but it is beyond the accuracy of our experimental
data.

In this model there is no mixing of the basis functions of $\vec{k}_1$ and
$\vec{k}_2$ on the same orbit and thus it gives constant moment
configurations for any direction of the spins. The assumption of
having the same mixing coefficient for the atoms on the different
orbits and belonging to the different arms is not dictated by
symmetry, because the coefficients $C_{\lambda,\vec{k}_L}$ in formula
(\ref{Sofbf}) are independent quantities for $\vec{k}_1$ and $\vec{k}_2$. However, in our particular case this assumption gives an excellent fit and good spin values as shown in Table~\ref{magmom} (Model B).

The best fit magnetic configuration is shown in Fig.~\ref{mstr}. The
figure shows 1/4 part of the magnetic unit cell. The whole magnetic
cell contains 48 magnetic atoms. The mutual orientation of the spins
in the trimers are different for the trimers on orbits 1 and 2. On
the 1st orbit the Ni and Cu spins are close to antiparallel
configuration, whereas on the 2nd orbit they are close to a
ferromagnetic coupling. Intuitively one would expect to have the
same spin orientation in all the trimers, because the intra-trimer
interactions should be the strongest ones. To make the trimers
identical one should constrain the Ni-spin to be parallel to
$c$-axis and the Cu-spin to be in the plane perpendicular to the
c-axis. For this configuration Cu and Ni spins are perpendicular for
both trimers. However, this constrained model gives a very bad
goodness of fit $\chi^2=8.67$ and has to be rejected.  Since the
spins in the trimers are close to a parallel orientation we also
tried to constrain them to be parallel (Model C in
Table~\ref{magmom}). This constrained model gives only slightly
worse $\chi^2$ than the general case. The model contains only 4
adjustable parameters: two spin values and two angles that is not
bad for describing the intensities of 46 magnetic Bragg peaks.  The
spins are aligned roughly along $c$-axis with very small canting as
shown in Table~\ref{magmom}. We note that the sub-lattices of the
trimers corresponding to the different arms of the star
\{$\vec{k}$\} do not interfere with each other and the magnetic
configuration, in which all the spins in the trimers on the same orbit are
reversed will give the same Bragg peak intensities.  Thus we can have two type of domains with the reversed mutual orientation trimer spins on orbit 1 and orbit 2. 

From the above proof of the equivalence of the structure factors for the orbit 1 with $\vec{k}_1$ and the orbit 2 with $\vec{k}_2$ one can see that the model with the atoms from orbit 2 and with the propagation vector $\vec{k}_1$ (``orbit2+k1'') is not an allowed equivalent solution. We tried to find a solution by fitting the data to the ``orbit2+k1'' model, but the best solution has much worse $\chi^2_{\rm Bragg}=5.1$ similar to the Model A. The reason is the difference of the phase factors for Ni in the structure factor $\vec{F}(\vec{H})$. For example, the phase factor for Ni-atoms in the structure factor for the $({1\over2},0,0)$ peak is proportional to $\sin(\pi x)$ for orbit 1, but to $\sin(\pi y)$ for orbit 2 (we use the primitive cell settings). It is clear that the $x$ and $y$-coordinates have no any symmetry relation with the spin value $M_\perp$. Thus the intensity of the $({1\over2},0,0)$ peak for the orbit 2 and propagation vector $\vec{k}_1$ will depend also on the particular value of $y$, but not only on the $M_\perp$. 

\section{Discussion and conclusions}
\label{discon}

The $\rm Ni^{2+}$ ion always likes to occupy the end positions of
the trimers.  The valence of Ni if it would occupy the position Cu1
in the center of the trimer have reasonable value according to BVS
calculations (Table \ref{bvstab}), and so in this respect Ni could
occupy the middle position. Apparently the total crystal energy is
minimized for the Cu being in the middle position. It is interesting
to note that the distorted square pyramids $\rm Cu2O_5$ and the
distorted square planes $\rm Cu1O_4$  become much less distorted when going from (x=0) to (x=2) compositions (Table \ref{bvstab}). 
We note, that the decrease in the distortion of the polyhedron around Cu2-site in the x=2 compound where it is occupied solely by Ni-ions is in accordance with the fact that $\rm Ni^{2+}$ ($3d^8$) is non-Jahn-Teller active ion.

The trimers
in the (x=1) composition can be of three types Cu-Cu-Cu, Cu-Cu-Ni
and Ni-Cu-Ni with the statistical populations 25/50/25\%,
respectively. From the diffraction data alone we cannot determine
these populations, but from the analysis of the magnetic excitations
in the trimers \cite{furrer07} the real populations were determined
to be 36/28/36\%, implying that the non-symmetric Cu-Cu-Ni trimer is
significantly less populated with respect to the "ideal" statistical
value. The fact that Ni atom does not occupy the middle position has
precluded from the realization of the Bose-Einstein condensation in
this trimer system, however it might be worth trying to make the
Ni-substitution in the similar compounds \acu\ (A=Pb, Sr).

The antiferromagnetic ordering, which we have found in the (x=2)
sample occurs at much higher temperature ($T_N=20$~K) than in the
parent (x=0) compound ($T_N=0.9$~K). The higher $T_N$ might be due
to the increase in the dipole interaction strength: Ni-ion has two
times larger spin value and the intertrimer distances between
the Cu2-sites along $a$-axis and $z$-axis are decreased from
4.8~\AA\ to 4.5~\AA, and from 6.5~\AA\ to 6.4~\AA, respectively. In
addition, the superexchange (SE) coupling is expected to be larger
in (x=2) compound. The SE interaction between the (1D) trimer chains
along $b$-axis (as explained in Sec~\ref{cryst_sec}) is mediated by
the Cu2-O-P2-O-Cu2 path. The completely closed $\rm P^{5+}$
2p-shells provide SE path that can be both anti- and ferromagnetic.
The average Cu-O-P bond angle is increased from $126.3^\circ$ to
$128.5^\circ$, implying that the antiferromagnetic SE via
Ni(3d)-O(2p) orbitals can be larger in the (x=2) compound.

The antiferromagnetic structure can be well described only using
both arms of the propagation vector star. It is quite unusual that
the two-k case can be revealed from the unpolarized powder
diffraction data analysis. However, the two-k solution excellently
describes the data with minimal number of the refined parameters.
For further verification of our magnetic structure model the single
crystal diffraction experiments might be useful. According to our
model the spin orientation in the Ni-Cu-Ni trimers can be both anti-
and ferromagnetic. The chains of the trimers running along the
$a$-axis are of two types, one consisting of the antiferromagnetic
(AFM) trimers and another one with the ferromagnetic (FM) trimers.
The $\rm Ni^{2+}$ ordered magnetic moment 1.9~$\mu_B$ is close to
the saturated value, whereas the $\rm Cu^{2+}$ moment 0.6~$\mu_B$ is
substantially smaller than the spin-only value. This might be due to
frustration of the Cu-moment, i.e. some trimers in the AFM trimer
chain have Cu-spins aligned ferromagnetically with Ni ones and vice
versa for the FM-chains.

In conclusion, we have successfully synthesized and studied the
crystal and magnetic structures of the novel mixed spin trimers
\cacuni\ (x=0,1,2) by means of neutron diffraction in the
temperature range 1.5-290~K. Our work forms an important ground for
the inelastic neutron scattering study of the dynamic magnetic
properties of this system.

\section*{A{\lowercase{cknowledgements}}}

This study was performed at Swiss neutron spallation SINQ and Swiss
light source SLS of Paul Scherrer Institute PSI (Villigen, PSI). We
thank L.~Keller for the help in neutron diffraction measurements and
P.~Fischer, O.~Zaharko and Zoso L. Davies for the discussions. Financial support by
the NCCR MaNEP project is gratefully acknowledged.

\bibliography{../../../refs/refs_general,../cacuni,../../../publication_list/publist2007}

\begin{thebibliography}{15}
\expandafter\ifx\csname natexlab\endcsname\relax\def\natexlab#1{#1}\fi
\expandafter\ifx\csname bibnamefont\endcsname\relax
  \def\bibnamefont#1{#1}\fi
\expandafter\ifx\csname bibfnamefont\endcsname\relax
  \def\bibfnamefont#1{#1}\fi
\expandafter\ifx\csname citenamefont\endcsname\relax
  \def\citenamefont#1{#1}\fi
\expandafter\ifx\csname url\endcsname\relax
  \def\url#1{\texttt{#1}}\fi
\expandafter\ifx\csname urlprefix\endcsname\relax\def\urlprefix{URL }\fi
\providecommand{\bibinfo}[2]{#2}
\providecommand{\eprint}[2][]{\url{#2}}

\bibitem[{\citenamefont{Belik et~al.}(2002)\citenamefont{Belik, Malakhov,
  Lazoryak, and Khasanov}}]{belik02}
\bibinfo{author}{\bibfnamefont{A.~A.} \bibnamefont{Belik}},
  \bibinfo{author}{\bibfnamefont{A.~P.} \bibnamefont{Malakhov}},
  \bibinfo{author}{\bibfnamefont{B.~I.} \bibnamefont{Lazoryak}},
  \bibnamefont{and} \bibinfo{author}{\bibfnamefont{S.~S.}
  \bibnamefont{Khasanov}}, \bibinfo{journal}{J. Solid State Chem.}
  \textbf{\bibinfo{volume}{163}}, \bibinfo{pages}{121} (\bibinfo{year}{2002}).

\bibitem[{\citenamefont{Matsuda et~al.}(2005)\citenamefont{Matsuda, Kakurai,
  Belik, Azuma, Takano, and Fujita}}]{matsuda05}
\bibinfo{author}{\bibfnamefont{M.}~\bibnamefont{Matsuda}},
  \bibinfo{author}{\bibfnamefont{K.}~\bibnamefont{Kakurai}},
  \bibinfo{author}{\bibfnamefont{A.~A.} \bibnamefont{Belik}},
  \bibinfo{author}{\bibfnamefont{M.}~\bibnamefont{Azuma}},
  \bibinfo{author}{\bibfnamefont{M.}~\bibnamefont{Takano}}, \bibnamefont{and}
  \bibinfo{author}{\bibfnamefont{M.}~\bibnamefont{Fujita}},
  \bibinfo{journal}{Phys. Rev. B} \textbf{\bibinfo{volume}{71}},
  \bibinfo{pages}{144411} (\bibinfo{year}{2005}).

\bibitem[{\citenamefont{Drillon et~al.}(1993)\citenamefont{Drillon, Belaiche,
  Legoll, Aride, Boukhari, and Moqine}}]{drillon93}
\bibinfo{author}{\bibfnamefont{M.}~\bibnamefont{Drillon}},
  \bibinfo{author}{\bibfnamefont{M.}~\bibnamefont{Belaiche}},
  \bibinfo{author}{\bibfnamefont{P.}~\bibnamefont{Legoll}},
  \bibinfo{author}{\bibfnamefont{J.}~\bibnamefont{Aride}},
  \bibinfo{author}{\bibfnamefont{A.}~\bibnamefont{Boukhari}}, \bibnamefont{and}
  \bibinfo{author}{\bibfnamefont{A.}~\bibnamefont{Moqine}},
  \bibinfo{journal}{J. Mag. Mag. Mater.} \textbf{\bibinfo{volume}{128}},
  \bibinfo{pages}{83} (\bibinfo{year}{1993}).

\bibitem[{\citenamefont{Belik et~al.}(2005)\citenamefont{Belik, Matsuo, Azuma,
  Kindo, and Takano}}]{belik05}
\bibinfo{author}{\bibfnamefont{A.~A.} \bibnamefont{Belik}},
  \bibinfo{author}{\bibfnamefont{A.}~\bibnamefont{Matsuo}},
  \bibinfo{author}{\bibfnamefont{M.}~\bibnamefont{Azuma}},
  \bibinfo{author}{\bibfnamefont{K.}~\bibnamefont{Kindo}}, \bibnamefont{and}
  \bibinfo{author}{\bibfnamefont{M.}~\bibnamefont{Takano}},
  \bibinfo{journal}{J. Solid State Chem.} \textbf{\bibinfo{volume}{178}},
  \bibinfo{pages}{709} (\bibinfo{year}{2005}).

\bibitem[{\citenamefont{Ruegg et~al.}(2003)\citenamefont{Ruegg, Cavadini,
  Furrer, Gudel, Kramer, Mutka, Habicht, Vorderwisch, and Wildes}}]{ruegg03}
\bibinfo{author}{\bibfnamefont{C.}~\bibnamefont{Ruegg}},
  \bibinfo{author}{\bibfnamefont{N.}~\bibnamefont{Cavadini}},
  \bibinfo{author}{\bibfnamefont{A.}~\bibnamefont{Furrer}},
  \bibinfo{author}{\bibfnamefont{H.~U.} \bibnamefont{Gudel}},
  \bibinfo{author}{\bibfnamefont{K.}~\bibnamefont{Kramer}},
  \bibinfo{author}{\bibfnamefont{H.}~\bibnamefont{Mutka}},
  \bibinfo{author}{\bibfnamefont{A.~K.} \bibnamefont{Habicht}},
  \bibinfo{author}{\bibfnamefont{P.}~\bibnamefont{Vorderwisch}},
  \bibnamefont{and} \bibinfo{author}{\bibfnamefont{A.}~\bibnamefont{Wildes}},
  \bibinfo{journal}{Nature} \textbf{\bibinfo{volume}{423}}, \bibinfo{pages}{62}
  (\bibinfo{year}{2003}).

\bibitem[{\citenamefont{Podlesnyak et~al.}(2007)\citenamefont{Podlesnyak,
  Pomjakushin, Pomjakushina, Conder, and Furrer}}]{furrer07}
\bibinfo{author}{\bibfnamefont{A.}~\bibnamefont{Podlesnyak}},
  \bibinfo{author}{\bibfnamefont{V.}~\bibnamefont{Pomjakushin}},
  \bibinfo{author}{\bibfnamefont{E.}~\bibnamefont{Pomjakushina}},
  \bibinfo{author}{\bibfnamefont{K.}~\bibnamefont{Conder}}, \bibnamefont{and}
  \bibinfo{author}{\bibfnamefont{A.}~\bibnamefont{Furrer}},
  \bibinfo{journal}{Phys. Rev. B} \textbf{\bibinfo{volume}{76}},
  \bibinfo{pages}{064420} (\bibinfo{year}{2007}).

\bibitem[{\citenamefont{Fischer
  et~al.}(2000{\natexlab{a}})\citenamefont{Fischer, Frey, Koch, Koennecke,
  Pomjakushin, Schefer, Thut, Schlumpf, Buerge, Greuter et~al.}}]{hrpt}
\bibinfo{author}{\bibfnamefont{P.}~\bibnamefont{Fischer}},
  \bibinfo{author}{\bibfnamefont{G.}~\bibnamefont{Frey}},
  \bibinfo{author}{\bibfnamefont{M.}~\bibnamefont{Koch}},
  \bibinfo{author}{\bibfnamefont{M.}~\bibnamefont{Koennecke}},
  \bibinfo{author}{\bibfnamefont{V.}~\bibnamefont{Pomjakushin}},
  \bibinfo{author}{\bibfnamefont{J.}~\bibnamefont{Schefer}},
  \bibinfo{author}{\bibfnamefont{R.}~\bibnamefont{Thut}},
  \bibinfo{author}{\bibfnamefont{N.}~\bibnamefont{Schlumpf}},
  \bibinfo{author}{\bibfnamefont{R.}~\bibnamefont{Buerge}},
  \bibinfo{author}{\bibfnamefont{U.}~\bibnamefont{Greuter}},
  \bibnamefont{et~al.}, \bibinfo{journal}{Physica B}
  \textbf{\bibinfo{volume}{276-278}}, \bibinfo{pages}{146}
  (\bibinfo{year}{2000}{\natexlab{a}}).

\bibitem[{\citenamefont{Fischer
  et~al.}(2000{\natexlab{b}})\citenamefont{Fischer, Keller, Schefer, and
  Kohlbrecher}}]{dmc}
\bibinfo{author}{\bibfnamefont{P.}~\bibnamefont{Fischer}},
  \bibinfo{author}{\bibfnamefont{L.}~\bibnamefont{Keller}},
  \bibinfo{author}{\bibfnamefont{J.}~\bibnamefont{Schefer}}, \bibnamefont{and}
  \bibinfo{author}{\bibfnamefont{J.}~\bibnamefont{Kohlbrecher}},
  \bibinfo{journal}{Neutron News} \textbf{\bibinfo{volume}{11}},
  \bibinfo{pages}{19} (\bibinfo{year}{2000}{\natexlab{b}}).

\bibitem[{\citenamefont{Rodriguez-Carvajal}(1993)}]{Fullprof}
\bibinfo{author}{\bibfnamefont{J.}~\bibnamefont{Rodriguez-Carvajal}},
  \bibinfo{journal}{Physica B} \textbf{\bibinfo{volume}{192}},
  \bibinfo{pages}{55} (\bibinfo{year}{1993}).

\bibitem[{\citenamefont{Lazoryak et~al.}(1999)\citenamefont{Lazoryak, Khan,
  Morozov, Belik, and Khasanov}}]{lazoryak99}
\bibinfo{author}{\bibfnamefont{B.~I.} \bibnamefont{Lazoryak}},
  \bibinfo{author}{\bibfnamefont{N.}~\bibnamefont{Khan}},
  \bibinfo{author}{\bibfnamefont{V.~A.} \bibnamefont{Morozov}},
  \bibinfo{author}{\bibfnamefont{A.~A.} \bibnamefont{Belik}}, \bibnamefont{and}
  \bibinfo{author}{\bibfnamefont{S.~S.} \bibnamefont{Khasanov}},
  \bibinfo{journal}{J. Solid State Chem.} \textbf{\bibinfo{volume}{145}},
  \bibinfo{pages}{345} (\bibinfo{year}{1999}).

\bibitem[{\citenamefont{Favre-Nicolin and Cerny}(2002)}]{fox}
\bibinfo{author}{\bibfnamefont{V.}~\bibnamefont{Favre-Nicolin}}
  \bibnamefont{and} \bibinfo{author}{\bibfnamefont{R.}~\bibnamefont{Cerny}},
  \bibinfo{journal}{J. Appl. Cryst.} \textbf{\bibinfo{volume}{35}},
  \bibinfo{pages}{734} (\bibinfo{year}{2002}).

\bibitem[{\citenamefont{Brown and Altermatt}(1985)}]{brown85}
\bibinfo{author}{\bibfnamefont{I.~D.} \bibnamefont{Brown}} \bibnamefont{and}
  \bibinfo{author}{\bibfnamefont{D.}~\bibnamefont{Altermatt}},
  \bibinfo{journal}{Acta Cryst. B} \textbf{\bibinfo{volume}{41}},
  \bibinfo{pages}{244} (\bibinfo{year}{1985}).

\bibitem[{\citenamefont{Izyumov et~al.}(1991)\citenamefont{Izyumov, Naish, and
  Ozerov}}]{izyumov91}
\bibinfo{author}{\bibfnamefont{Y.~A.} \bibnamefont{Izyumov}},
  \bibinfo{author}{\bibfnamefont{V.~E.} \bibnamefont{Naish}}, \bibnamefont{and}
  \bibinfo{author}{\bibfnamefont{R.~P.} \bibnamefont{Ozerov}},
  \emph{\bibinfo{title}{Neutron diffraction of magnetic materials}}
  (\bibinfo{publisher}{New York [etc.]: Consultants Bureau},
  \bibinfo{year}{1991}).

\bibitem[{\citenamefont{Kovalev}(1993)}]{kovalev}
\bibinfo{author}{\bibfnamefont{O.~V.} \bibnamefont{Kovalev}},
  \emph{\bibinfo{title}{Representations of the Crystallographic Space Groups:
  irreducible representations, induced representations, and corepresentations}}
  (\bibinfo{publisher}{Gordon and Breach Science Publishers},
  \bibinfo{year}{1993}), \bibinfo{edition}{2nd} ed.

\bibitem[{\citenamefont{Kirkpatrick et~al.}(1983)\citenamefont{Kirkpatrick,
  Gelatt, and Vecchi}}]{kirkpatrick83}
\bibinfo{author}{\bibfnamefont{S.}~\bibnamefont{Kirkpatrick}},
  \bibinfo{author}{\bibfnamefont{C.~D.} \bibnamefont{Gelatt}},
  \bibnamefont{and} \bibinfo{author}{\bibfnamefont{M.~P.}
  \bibnamefont{Vecchi}}, \bibinfo{journal}{Science}
  \textbf{\bibinfo{volume}{220}}, \bibinfo{pages}{671} (\bibinfo{year}{1983}).

\end{thebibliography}


\newpage
\setlength{\LTcapwidth}{\textwidth} \setlength\LTleft{-10pt}
\setlength\LTright{0pt} \setlongtables
\begin{longtable}{l|l|l|l}

\caption{The structure parameters and Cu(Ni)-O interatomic
  distances
  in $\rm Ca_3Cu_3(PO_4)_4 $ (x=0), and $\rm Ca_3Cu_2Ni(PO_4)_4 $
  (x=1) [sp.gr. $P\,1\,2_1/a\,1$ (no. 14)] and and $\rm Ca_3CuNi_2(PO_4)_4$
  (x=2) [sp.gr. $C\,1\,2/c  \,1$ (no. 15)].  The Wyckoff positions are
  (2a) for Cu1, (2b) for Ca1 and (4e) for other atoms for the
  compounds with (x=0,1) and (4b) for Cu1, (4e) for Ca1 and (8f) for
  other atoms for the (x=2) one.  The data are refined from the
  powder neutron
  diffraction patterns measured at HRPT/SINQ with wavelength
  $\lambda=1.886$~\AA.  Bragg reliability factor $R_{\rm Bragg}$ for
  the main phase and conventional reliability factors for the whole
  pattern $R_{\rm wp}, R_{\rm exp}, \chi^2$ are also given. The notation of
  the oxygen atoms around the Cu2-Cu1-Cu2 trimer are given in
  Fig.~\ref{trimer}. The bond lengths are given in \AA, the angles in
  degrees and the isotropic thermal displacement parameter $B$ in
  $\rm\AA^2$. For the Cu1 and Cu2 positions the refined occupancies
  Cu/Ni are given.}\\

\label{tab1}


                           &x=0                               &x=1                               &x=2 \\
\hline
$a$, \AA                     &17.62154(8)                       &17.71388(9)                       &17.7174(1)                         \\
$b$, \AA                     &4.90205(2)                        &4.88512(2)                        &4.82109(4)                         \\
$c$, \AA                     &8.92224(5)                        &8.84635(5)                        &17.8475(1)                         \\
$\gamma$, deg                &124.0744(3)                       &123.8436(3)                       &123.6373(5)                        \\
$V$, $\AA^3$                 &638.39                            &635.81                            &1269.22                           \\
\hline
$x,y,z$                    &$0,0,0$                           &$0,0,0$                           &${1\over2},0,0$                      \\
$B$          (Cu1)         &0.64(4)                           &0.74(6)                           &0.81(8)                              \\
     Cu/Ni                 &1/0                               &0.980/0.020(15)                   &0.84/0.16(2)                         \\
\hline
$x,y,z$                    &0.1198(1),0.4771(3),0.9430(2)   &0.1213(1),0.4717(3),0.9461(2) &0.6199(1),0.4638(4),0.4671(1)\\
$B$          (Cu2)         &0.39(4)                           &0.70(3)                           &0.67(4)                               \\
     Cu/Ni                 &1/0                               &0.510/0.490(8)                    &0.08/0.92(1)                           \\
\hline
$x,y,z$                    &${1\over2},0,{1\over2}$           &${1\over2},0,{1\over2}$           &0,0.951(1),${1\over4}$           \\
$B$          (Ca1)         &0.44(6)                           &0.71(8)                           &0.42(9)                               \\
\hline
$x,y,z$                    &0.2659(2),0.4630(6),0.7265(3)   &0.2648(2),0.4648(6),0.7267(4) &0.7667(2),0.4441(8),0.3614(2)\\
$B$          (Ca2)         &0.80(5)                            &0.63(6)                           &0.69(7)\\
\hline
$x,y,z$                    &0.5934(1),0.9897(5),0.2487(3)   &0.5930(2),0.9857(7),0.2480(3) &0.0933(2),0.9915(7),0.1257(2)\\
$B$          (P1)          &0.39(4)                           &0.67(4)                           &0.22(5)\\
\hline
$x,y,z$                    &0.8406(2),0.0161(5),0.2164(3)  &0.8417(2),0.0175(5),0.2194(3) &0.3406(2),0.9498(7),0.1111(2)\\
$B$          (P2)          &0.40(4)                           &0.36(5)                           &0.35(6)\\
\hline
$x,y,z$                    &0.6794(2),0.9046(4),0.4240(3) &0.6798(2),0.9023(5),0.4267(3) &0.1793(2),0.0811(6),0.2121(2)\\
$B$          (O1)          &0.68(4)                           &0.68(4)                           &0.35(5)\\
\hline
$x,y,z$                    & 0.0087(1),0.5519(5),0.2534(3) &0.0098(1),0.5524(5),0.2536(3) &0.5111(2),0.5214(7),0.1301(2)\\
$B$          (O2)          &0.68(4)                           &0.61(6)                           &0.99(6)\\
\hline
$x,y,z$                    &0.6006(1),0.2866(4),0.2004(3) &0.5997(2),0.2810(5),0.1927(4) &0.1017(2),0.6968(6),0.0995(2)\\
$B$          (O3)          &0.73(4)                           &0.85(4)                           &0.68(6)\\
\hline
$x,y,z$                    &0.5812(1),0.8074(4),0.0918(3) &0.5846(2),0.7987(5),0.0962(4) &0.0801(2),0.1812(6),0.0477(2)\\
$B$          (O4)          &0.44(4)                           &0.76(5)                           &0.45(6)\\
\hline
$x,y,z$                    &0.8978(2),0.8549(4),0.3861(3) &0.9014(2),0.8629(5),0.3913(4) &0.3980(2),0.1037(5),0.1972(2)\\
$B$          (O5)          &1.09(4)                           &1.62(6)                           &2.04(8)\\
\hline
$x,y,z$                    &0.8542(1),0.9282(4),0.0661(3) &0.8518(2),0.9277(5),0.0641(3) &0.3563(2),0.0482(5),0.0381(2)\\
$B$          (O6)          &0.66(5)                           &1.07(5)                           &1.11(6)\\
\hline
$x,y,z$                    &0.3574(1),0.1752(4),0.2467(3) &0.3586(2),0.1658(5),0.2433(3) &0.8542(2),0.1343(6),0.1220(2)\\
$B$          (O7)          &0.82(4)                           &1.20(6)                           &1.01(6)\\
\hline
$x,y,z$                    &0.7396(1),0.9661(5),0.1521(3) &0.7429(2),0.9709(6),0.1622(3) &0.2412(2),0.9966(6),0.0806(2)\\
$B$          (O8)          &0.83(4)                           &0.95(5)                           &1.43(6)\\
\hline
$R_{\rm Bragg},\%$          &1.84                              &1.68                              &1.69                  \\
$R_{\rm wp},R_{\rm exp},\chi^2$   &      2.48, 0.92, 7.23            &      2.68, 1.11, 5.82            &      3.0 , 1.57, 3.68\\
\hline
Cu1-O4                     &1.917(2)                          &1.918(2)                          &1.939(3)      \\
Cu2-O4                     &2.086(4)                          &2.099(4)                          &2.045(4)  \\
Cu1-Cu2                    &3.534(2)                          &3.556(2)                          &3.352(2)  \\
Cu1-O4-Cu2                 &123.9(2)                        &124.5(2)                        &114.6(2)    \\
\hline
Cu1-O3                     &1.970(2)                        &1.953(2)                          &1.938(2)  \\
Cu2-O2                     &1.942(2)                          &1.989(2)                          &2.003(3)  \\
Cu2-O7                     &2.175(4)                          &2.023(3)                          &2.055(5)  \\
Cu2-O8                     &1.898(2)                          &2.121(4)                          &1.984(3)  \\
Cu2-O6                     &2.051(3)                          &1.946(2)                          &2.060(4)      \\

\end{longtable}
\newpage

\begin{table}

\caption{The average cation-oxygen bond lengths $d$, the rms
distortion of the polyhedra $\delta d/d$ in units $10^{-4}$ and the
bond valence sum BVS calculated from the experimental distances
using {\tt FULLPROF} \cite{Fullprof} suite and the BVS parameters
from \cite{brown85}. C and V are the coordination of the polyhedra
and the nominal cation valence, respectively.}
 \label{bvstab}

\begin{center}
\begin{tabular}{l|l|l|l|l|l|l|l|l|l|l|l}
\multicolumn{3}{c|}{ }&\multicolumn{3}{c|}{$x=0$}&\multicolumn{3}{c|}{$x=1$}&\multicolumn{3}{c}{$x=2$}\\
\hline
At. &C&V&$d$, \AA       &   $\delta d/d$ &  BVS & $d$, \AA       &   $\delta d/d$ &  BVS & $d$, \AA       &   $\delta d/d$ &  BVS \\
\hline
Cu1&4&2&$1.9450(   9)$&  1.5 &$ 1.953(  5)$&$ 1.9358(  10)$&  0.5&$ 2.000(  5)$&$ 1.9387(  13)$&0.002&$  1.983(  8)$\\
\hline
Ni1&4&2&$            $&      &$           $&$             $&     &$ 1.869(  5)$&$             $&     &$  1.853( 41)$\\
\hline
Cu2&5&2&$2.0290(  13)$& 24.3 &$ 2.013(  6)$&$ 2.0336(  13)$& 10.9&$ 1.950( 12)$&$             $&     &$  1.947( 41)$\\
\hline
Ni2&5&2&$            $&      &$           $&$             $&     &$ 1.822( 12)$&$ 2.0293(  16)$&  2.2&$  1.819(  8)$\\
\hline
Ca1&6&2&$2.3331(   9)$&  5.1 &$ 2.253(  6)$&$ 2.3273(   8)$&  4.4&$ 2.286(  5)$&$ 2.3361(  16)$& 18.4&$  2.290( 11)$\\
\hline
Ca2&3&2&$2.3714(  21)$&  2.0 &$ 1.010(  6)$&$ 2.3707(  22)$&  6.0&$ 1.020(  6)$&$ 2.4017(  25)$&  7.4&$  1.255(  8)$\\
\hline
P1 &4&5&$1.5384(  17)$&  2.7 &$ 4.958( 22)$&$ 1.5372(  18)$&  1.3&$ 4.969( 23)$&$ 1.5298(  23)$&  2.7&$  5.074( 31)$\\
\hline
P2 &4&5&$1.5343(  17)$&  2.5 &$ 5.014( 22)$&$ 1.5357(  18)$&  4.9&$ 5.005( 24)$&$ 1.5297(  23)$&  2.6&$  5.077( 32)$\\

\end{tabular}
\end{center}
\end{table}

\begin{table}

\caption{Positions of the magnetic atoms in the $\rm 0^{th}$ unit
cell,
  symmetry operators of $G_k$ and $G$ (space group $G=C 1 2/c 1$,
  propagation vector $\vec{k}=[{1\over2} {1\over2} 0]$) and basis
  functions for irreducible representation $\tau_2$. The Ni-atom is in
  general position (8f) with coordinates (0.62065,
  0.53530, 0.96795), the Cu-atom is in(4b) position (0, \mm{{1\over2}}, 0),
  The atoms of the orbit 2 are generated from the respective atoms of
  orbit 1 by the symmetry element $(2\,\, 0,y,{1\over4})$. The basis functions
  for orbit 2 are constructed using formula (\ref{karm}).}
 \label{magtab}

\begin{center}
\begin{tabular}{l|l|l|l|l|l}

Atom & eqv. pos.         & sym. op. $G; G_k$ & $\tau_2$ & $\tau_2'$
& $\tau_2''$\\\hline
\multicolumn{6}{l}{orbit 1,  $\vec{k}_1=[{1\over2} {1\over2} 0]$}\\
Ni11 & x,y,z             &  1;1                & 100      &  010      &  001 \\
Ni12 & -x+1,-y+1,-z+1    & $\bar{1}$;$\bar{1}$ & -100     &  0-10 &
00-1\\\hline Cu1  & 0,\mm{{1\over2}},0           &  1;1 & 100     &
010       &  001\\\hline
\multicolumn{6}{l}{orbit 2 (2 0,y,$1\over4$), $\vec{k}_2=[-{1\over2} {1\over2} 0]$}\\
Ni21 & -x+1,y,-z+$3\over4$     & 2 0,y,$1\over4$; 1        & 100      &  0-10     & 001 \\
Ni22 &  x-1,-y+1,z-\mm{{1\over2}}   & c x,0,z; -1         & -100 &
010     &  00-1\\\hline Cu2  & 0,\mm{{1\over2}},\mm{{1\over2}} & 2
0,y,$1\over4$; 1        & -100     &  010     &  00-1\\\hline

\end{tabular}
\end{center}
\end{table}

\begin{table}

\caption{Magnetic model parameters for \cacuni (x=2) refined from
the
  diffraction data shown in Fig.~\ref{difpatt}. The numeration of the
  atoms is the same as in Table~\ref{magtab}. $M$ is the size of the
  magnetic moment, $\phi$ and $\theta$ are spherical angles with $c$ (azimuth)
  and $b$ (zenith) axes, respectively. The graphical illustration of the spherical angles is given in Fig.~\ref{mstr}. The errorbars are given only for the
  independently refined parameters. The model A does not fit the data,
  but given for completeness. See the text for details. In the models B
  and C $\phi_{Ni21}=\phi_{Ni11}$, $\theta_{Ni21}=\pi-\theta_{Ni11}$,
  $\phi_{Cu2}=\phi_{Cu1}+\pi$, $\theta_{Cu2}=\theta_{Cu1}$. In the
  model C the spins of Ni and Cu are constrained to be (anti)parallel
  in the trimers.}

\label{magmom}

\begin{center}
\begin{tabular}{l|l|l|l|l|l|l|l|l|l}

     &  \multicolumn{3}{c|}{model A} & \multicolumn{3}{c|}{model B} &  \multicolumn{3}{c}{model C}  \\\hline
     &$M, \mu_B$       & $\phi$     & $\theta$ &$M, \mu_B$       & $\phi$     & $\theta$&$M, \mu_B$
& $\phi$     & $\theta$  \\

Ni11 &1.760(2) & 104.3(1.2) & 78.4(1.5)   &1.892(9) &   176.1(8) & 83.98(45)  &1.883(8)&  175.9(9)&   83.4(4) \\
Cu1  &1.196(4) & 21.2(4.4)  & 47.5(3.1)   &0.62(2)  & 153.5(3.7) & 103.3(2.9) &0.57(1) &  175.9   &   96.6     \\
Ni21 &1.760    &12.4(1.7)   & 69.9(1.3)   &1.892    & 176.1      & 96.02      &1.883   &  175.9   &   96.6    \\
Cu2  &1.196    &26.9(4.3)   & 127.7(4.1) &0.62     & 333.5      &
103.3      &0.565   &  355.9   &   96.6    \\\hline
\multicolumn{2}{l|}{$R_{wp}, R_{exp}, \chi^2, \chi^2_B$}&
\multicolumn{2}{l|}{14.1, 6.7, 4.4, 5.1} & \multicolumn{3}{l|}{8.9,
6.7, 1.76, 1.93}              &\multicolumn{3}{l}{9.2, 6.7, 1.86,
2.05} \\\hline

\end{tabular}
\end{center}
\end{table}

\def\extgra{pdf}
\def\figsiz{\textwidth}

\begin{figure}
  \begin{center}
    \includegraphics[width=\figsiz]{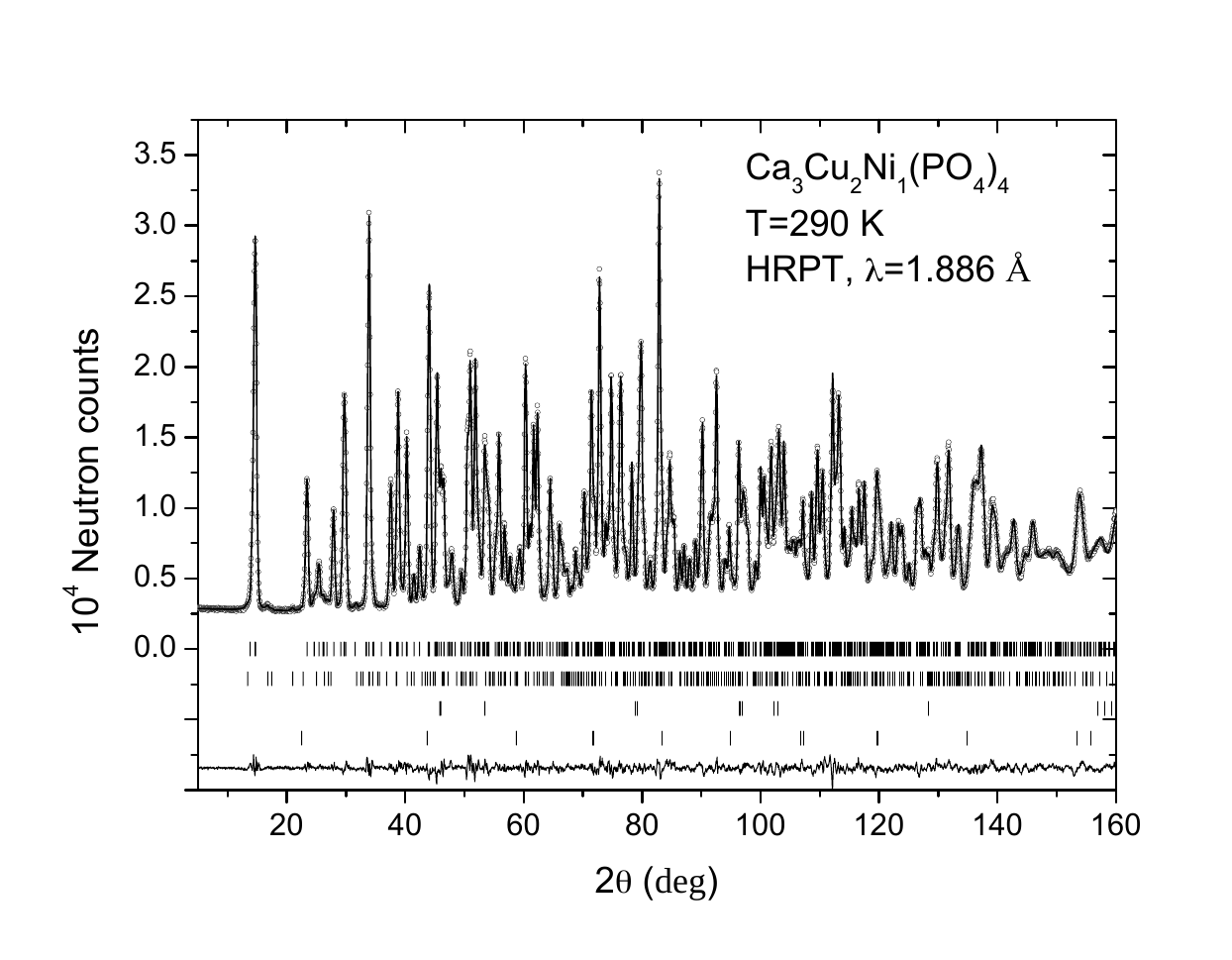} 
  \end{center}

  \caption{The Rietveld refinement pattern and difference plot of the
    neutron diffraction data for the sample \cacuni\ (x=1) at T=290K
    measured at HRPT with the wavelength $\lambda=1.886$~\AA. The rows
    of tics show the Bragg peak positions for the main phase, and two
    impurity phases: whitelockite, NiO nuclear and NiO magnetic peaks (from top to bottom).}
  \label{dpcu2ni}
\end{figure}

\begin{figure}
  \begin{center}
    \includegraphics[width=\figsiz]{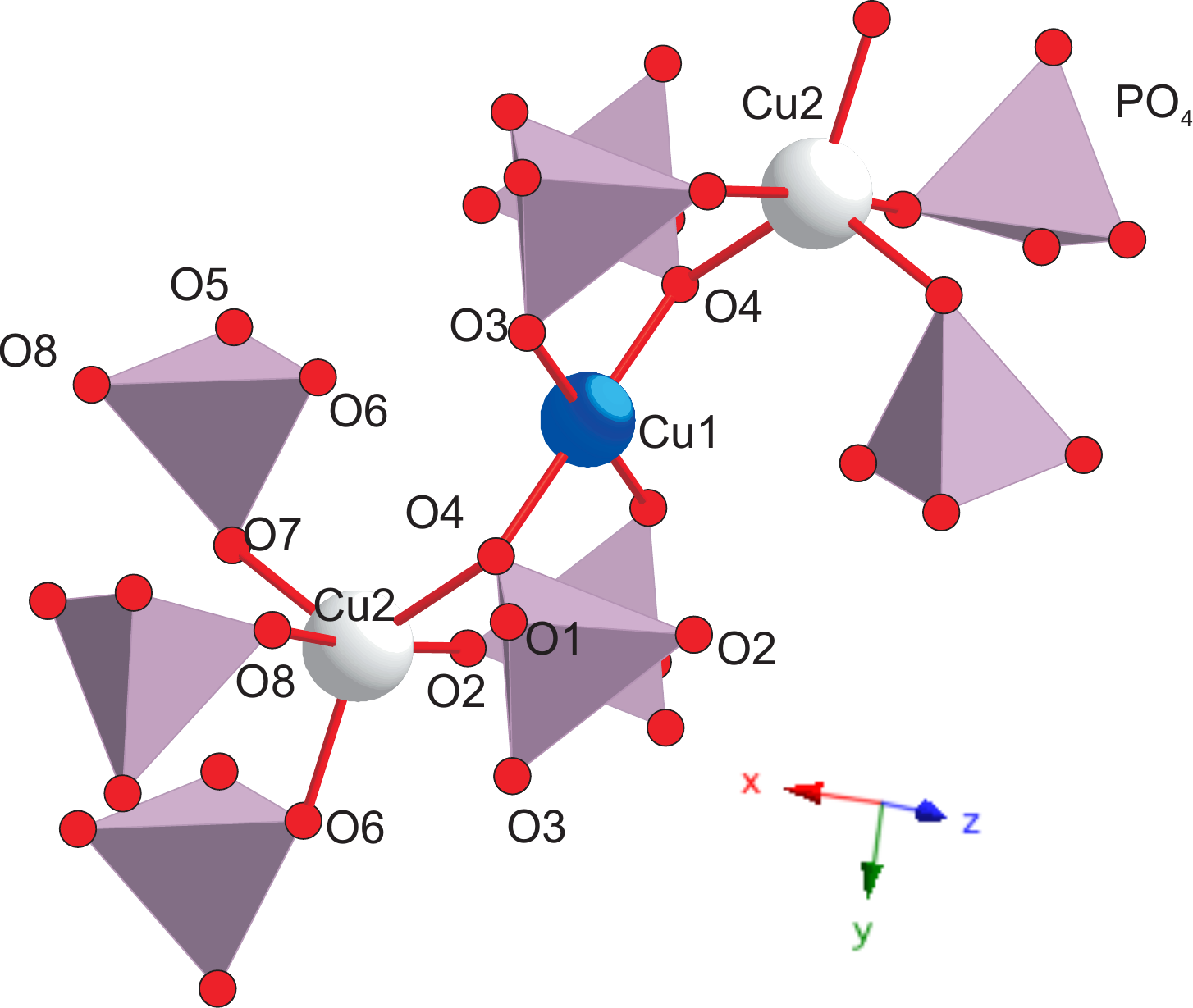} 
  \end{center}

  \caption{(Color online) Fragment of the crystal structure showing Cu2-O4-Cu1-O4-Cu2 trimer and
    the surrounding $\rm PO_4$ tetrahedra. The positions of the atoms and some selected interatomic distances
    are listed in Table ~\ref{tab1}.}
  \label{trimer}
\end{figure}

\begin{figure}
  \begin{center}
    \includegraphics[width=\figsiz]{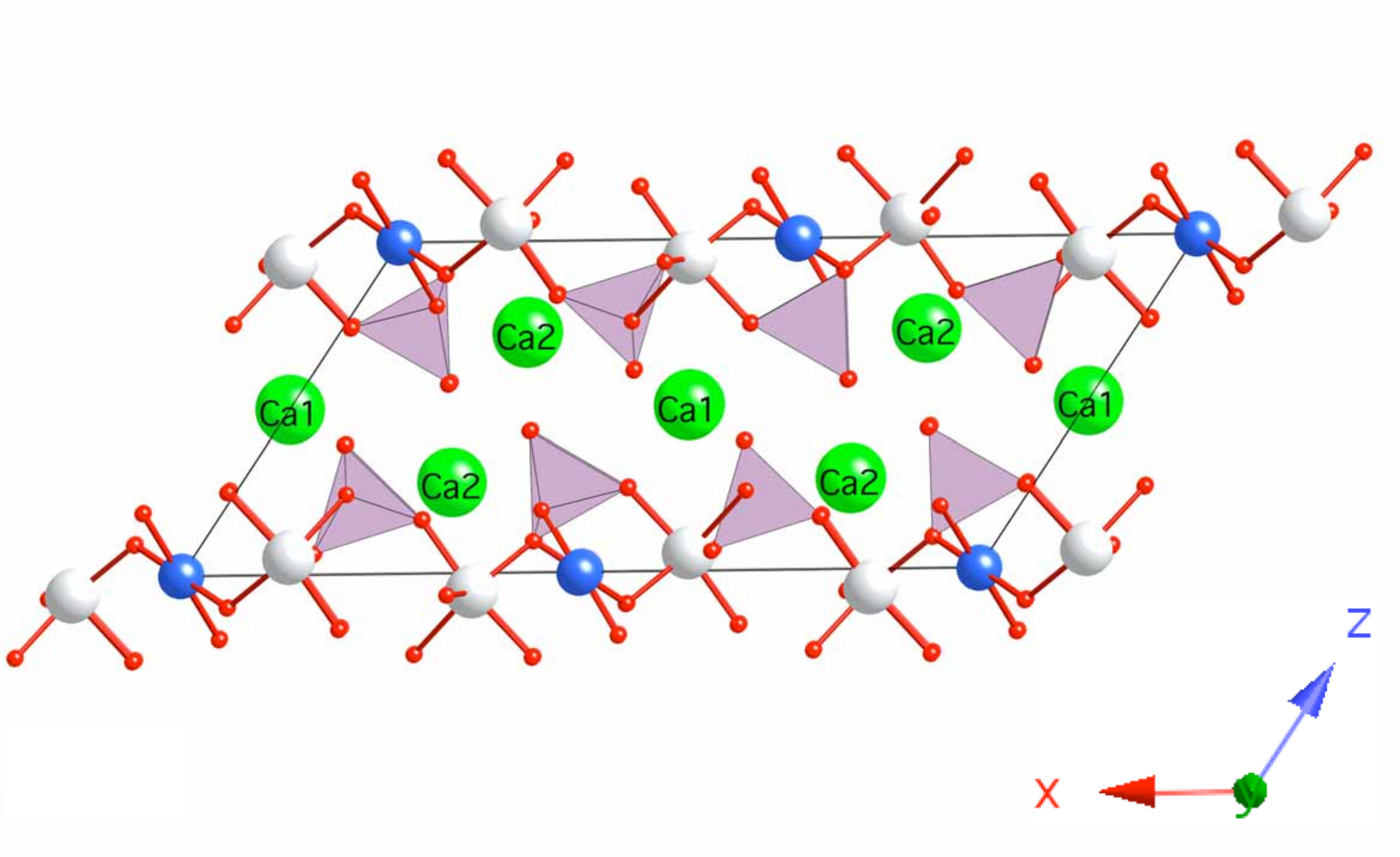} 
  \end{center}

  \caption{(Color online) Projection of the crystal structure of
  \cacuni\ on the $(ac)$ plane showing connectivity of the
  trimers. The Cu1 and Cu2 positions are shown by blue and white
  balls, the tetrahedra are $\rm PO_4$, the red sticks indicate Cu-O
  bonds, the big green balls are the Ca atoms.}
  \label{cstr}
\end{figure}

\begin{figure}
    \includegraphics[width=\figsiz]{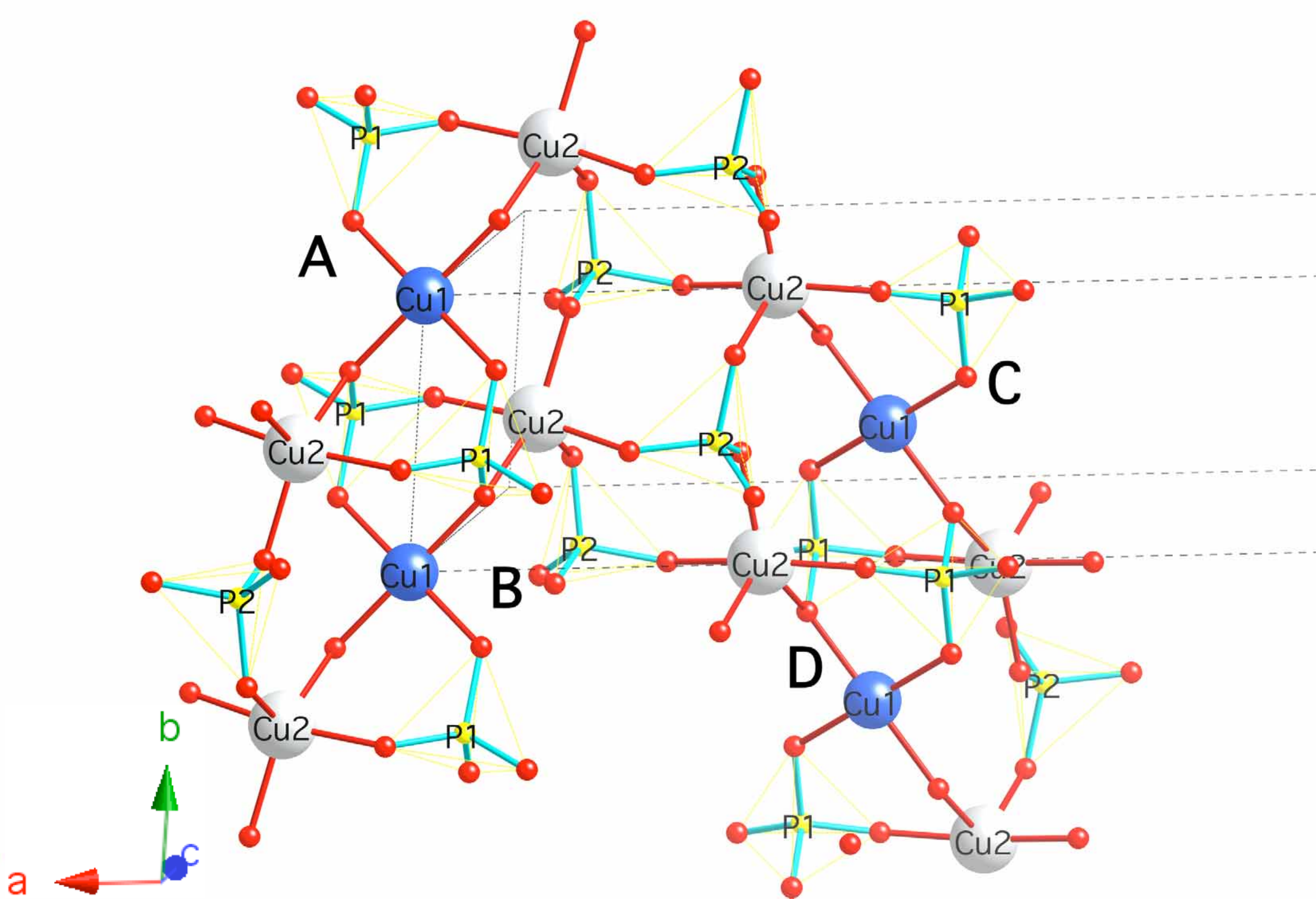} 

  \caption{(Color online) View of four trimers (A, B, C and D) projected approximately to the $(ab)$ plane
  in \cacuni\ showing possible superexchange paths between the
  trimers in $(ab)$ plane. The Cu1 and Cu2 positions are shown by blue and white
  balls, the tetrahedra with yellow balls in the center are $\rm PO_4$, the red sticks indicate Cu-O
  bonds, the Ca atoms are not shown. Each trimer is formed by the central Cu1 atom and two Cu2 atoms related by inversion with respect to the Cu1 position.}
  \label{trimer4}
\end{figure}

\begin{figure}
  \begin{center}
    \includegraphics[width=\figsiz]{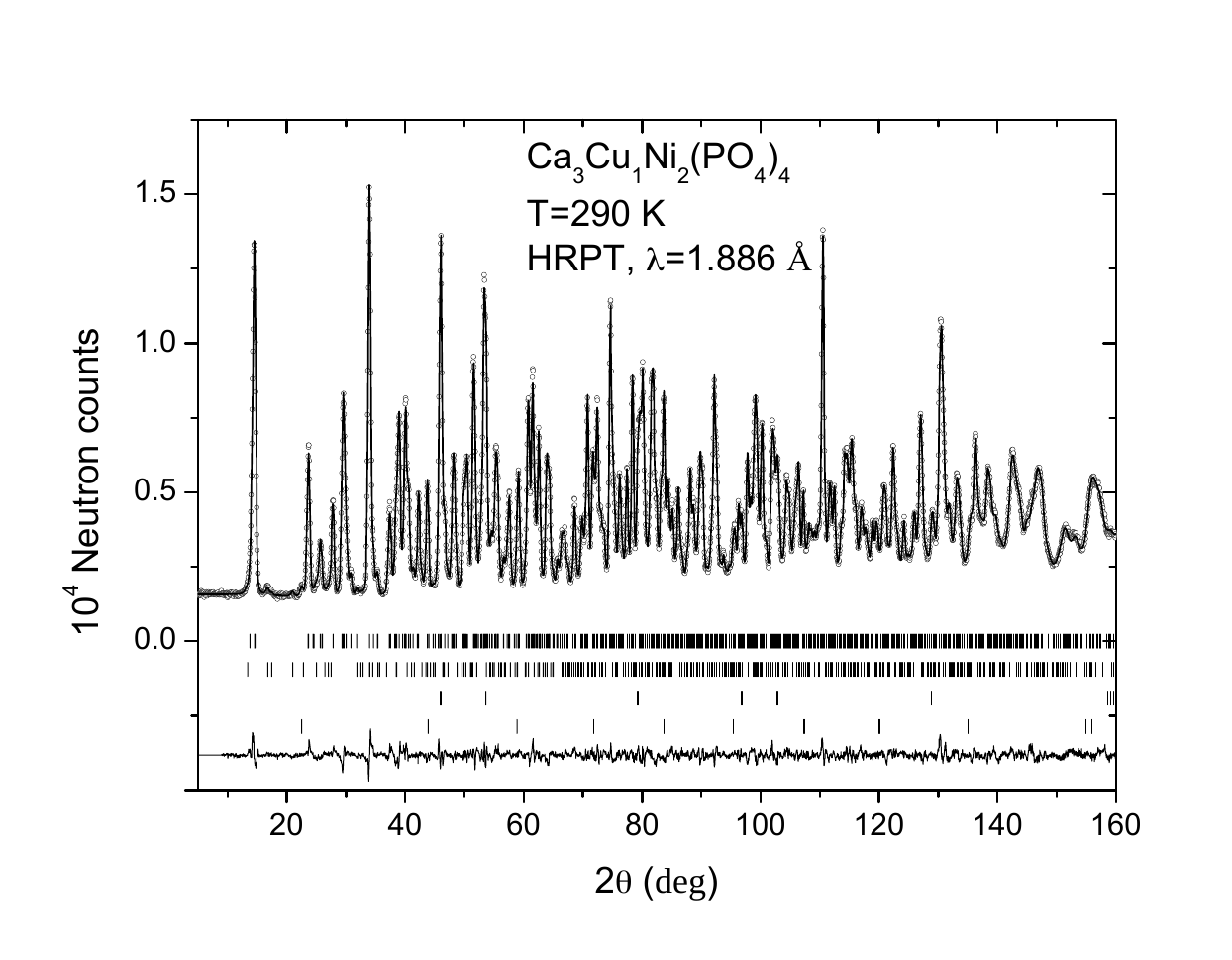} 
  \end{center}

  \caption{The Rietveld refinement pattern and difference plot of the
    neutron diffraction data for the sample \cacuni\ (x=2) at T=290K
    measured at HRPT with the wavelength $\lambda=1.886$~\AA. The rows
    of tics show the Bragg peak positions for the main phase and two
    impurity phases.}
  \label{dpcuni2}
\end{figure}

\begin{figure}
  \begin{center}
    \includegraphics[width=\figsiz]{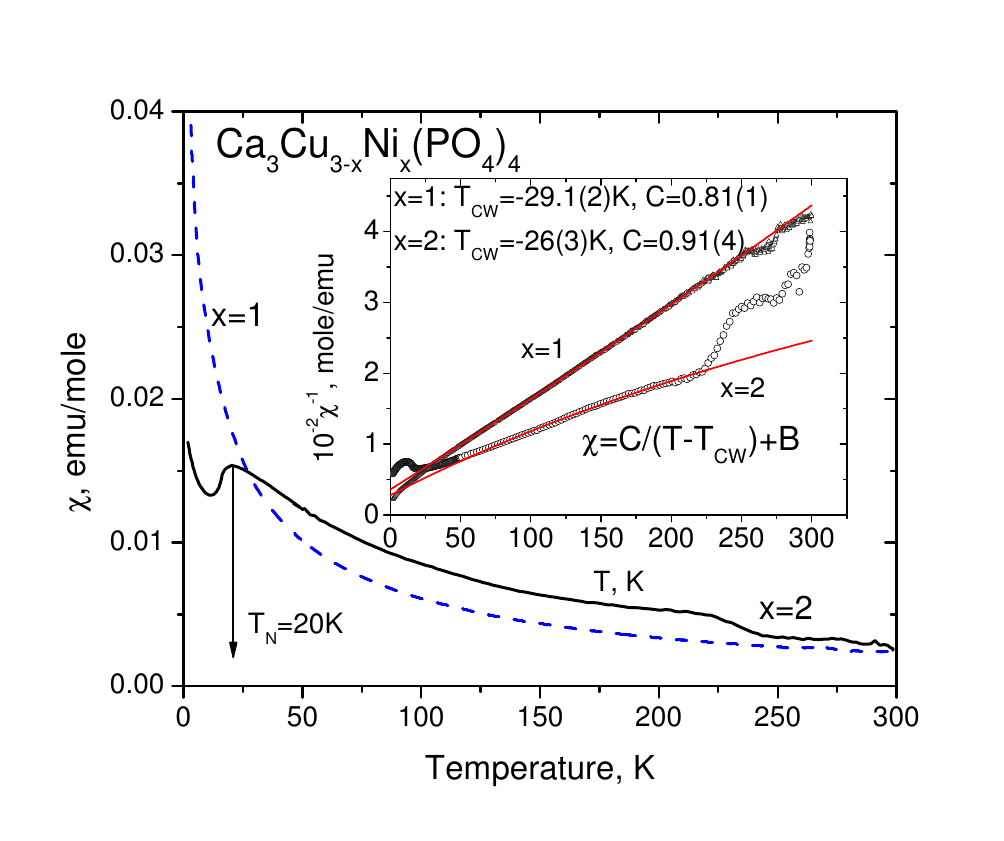} 
  \end{center}

  \caption{(Color online) Real part of the $ac$ magnetic
    susceptibility is shown as a function of temperature for the
    samples \cacuni\ with x=1 and 2. The inset shows the inverse
    susceptibility with the least-square fit to the Curie-Weiss law. The susceptibility is given per mole of the magnetic ions (Ni or Cu).
    The refined Curie-Weiss transition temperatures $T_{CW}$ and
    the Curie constants $C$ are show in inset.}
  \label{X(T)}
\end{figure}

\begin{figure}
  \begin{center}
    \includegraphics[width=\figsiz]{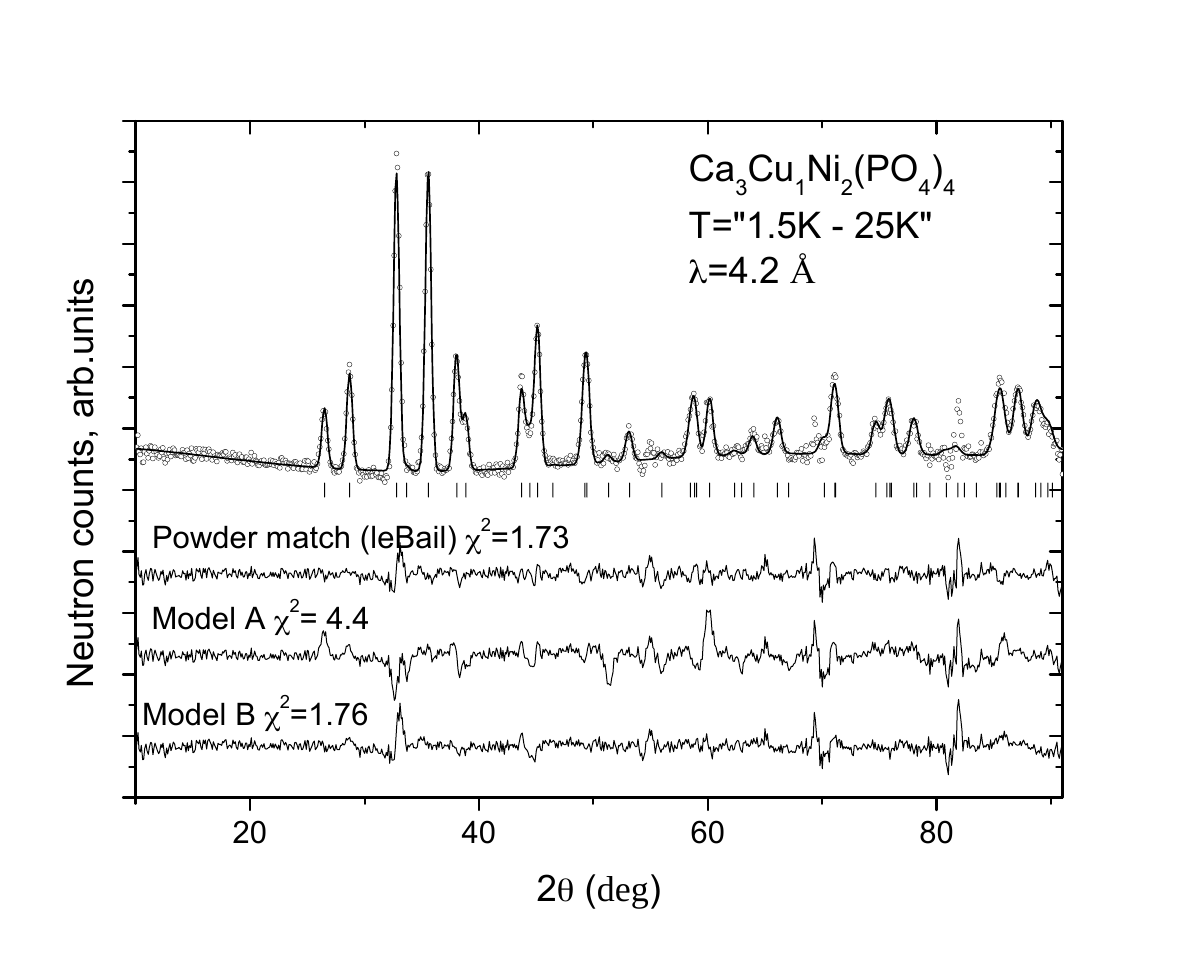} 
  \end{center}

  \caption{ The Rietveld refinement pattern and difference plot of the
    difference magnetic neutron diffraction pattern between 1.5K and
    25K for the sample with \cacuni\ (x=0.2) at T=290K at DMC with the
    wavelength $\lambda=4.2$~\AA. The difference curves are shown for
    3 different refinements: profile matching mode (the model contains
    only unit cell parameters and the propagation vector) and two
    different models. See the text for details.}
  \label{difpatt}
\end{figure}

\begin{figure}
  \begin{center}
    \includegraphics[width=\figsiz]{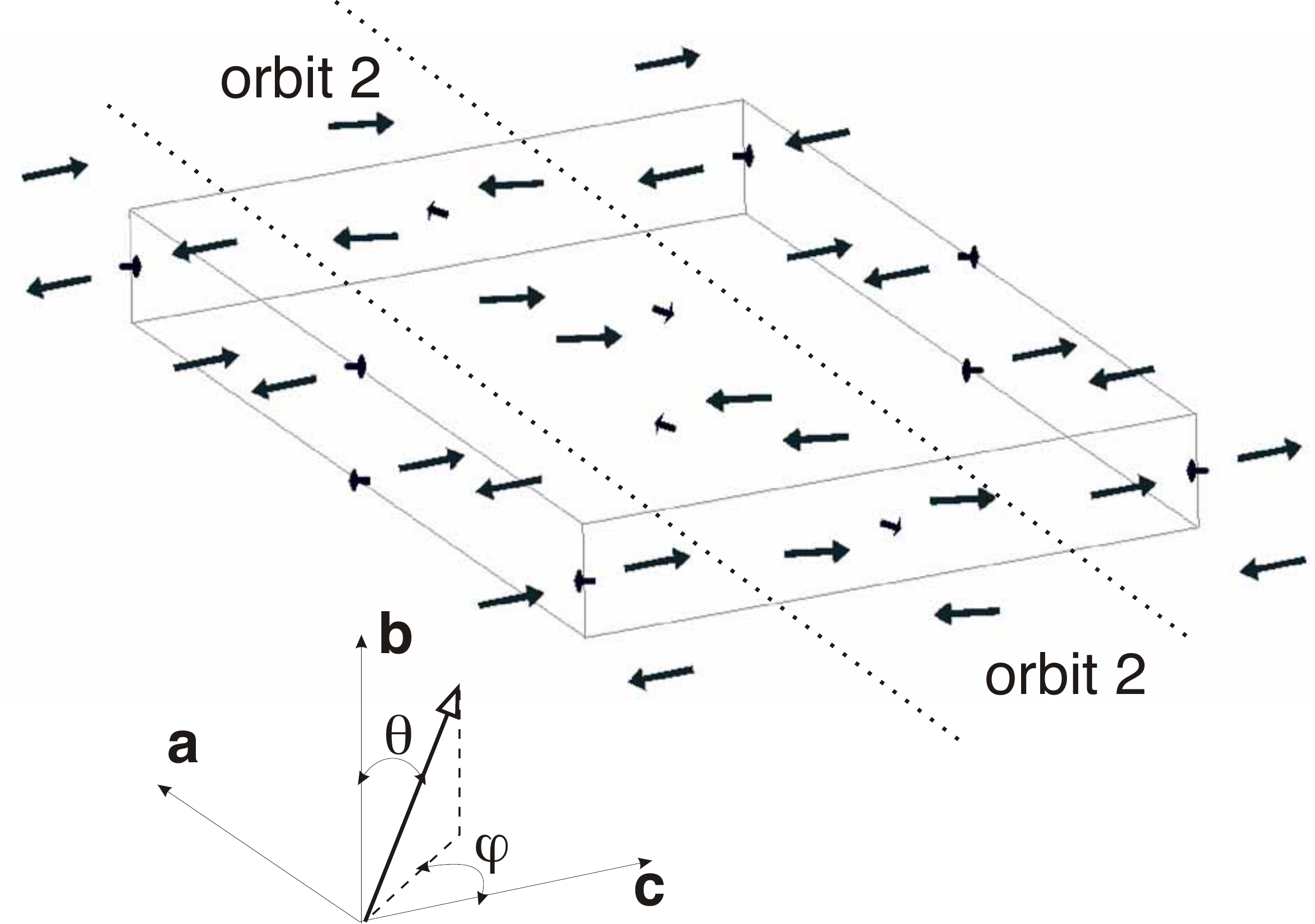}
  \end{center}

  \caption{The 0th unit cell of \cacuni\ (x=2) showing the
    configuration of the Ni and Cu spins. Some of the Ni-spins from
    the neighboring cells are shown for better visibility of the
    trimers. The unit cell constants at T=25~K are $a=17.724$~\AA,
    $b=4.815$~\AA, $c=17.836$~\AA, $\beta=123.756^\circ$ ($C2/c$ space
    group). The spins in the middle of the cell along $c$ direction between the dotted lines belong to the orbit 2 (Cu2, Ni21 and Ni22 spins) and 
    have propagation vector $\vec{k}_2=[-{1\over2} {1\over2} 0]$, the
    other spins belong to the orbit 1 (Cu1, Ni11 and Ni12 spins) and have propagation vector $\vec{k}_1=[{1\over2}
      {1\over2} 0]$. The structure corresponds to the model B shown in
    Table~\ref{magmom}. The crystal axes and the spherical angles used in Table~\ref{magmom} are also shown.}
  \label{mstr}
\end{figure}

\end{document}